\documentclass[aps,prl,preprint,groupedaddress,showpacs,floatfix]{revtex4}

\usepackage[]{graphicx}
\usepackage{epsfig}
\begin{document}


\title{Direct observation of interacting Kondo screened $4f$ moments in CePt$_5$ with XMCD}



\author{C.~Praetorius}
\author{A.~Koehl}
\altaffiliation{present address: Peter Gr{\"{u}}nberg Institut, Forschungszentrum J{\"{u}}lich, 52425 J{\"{u}}lich, Germany}
\author{B.~Muenzing}
\author{H.~Bardenhagen}
\author{K.~Fauth}
\email[]{fauth@physik.uni-wuerzburg.de}
\affiliation{Physikalisches Institut, Universit{\"{a}}t W{\"{u}}rzburg, Am Hubland, D--97074 W{\"{u}}rzburg}

\date{\today}

\begin{abstract}
We use x-ray absorption and magnetic circular dichroism to study electronic configuration and local susceptibility
 of CePt$_5$/Pt(111) surface alloys from well above to well below the impurity Kondo temperature. 
The anisotropic paramagnetic response is governed by the hexagonal crystal field and ferromagnetic
correlations with modified parameters for Ce moments residing next to the alloy surface.
Quantitative XMCD evaluations provide direct evidence of Kondo screening of both
spin and orbital $4f$ moments. 
Magnetic signatures of coherence are not apparent for $T \geq 13$ K.
\end{abstract}


\maketitle


Kondo and heavy fermion systems display fascinating many body phenomena
caused by the interplay of localized and itinerant electronic degrees of freedom. 
Despite their local character, hybridization and on site Coulomb repulsion 
give rise to unconventional macroscopic behavior and complex phase diagrams \cite{Hews93a,Loeh07a}.
Dilute magnetic impurities induce a many body singlet ground state, resulting
from effective antiferromagnetic interaction between localized and delocalized base states
\cite{Kond64a,Andr83a,Bick87a,Hews93a}.
Accordingly, a considerable density of excitations develops on
the scale of the singlet formation energy gain, frequently referred to as the Kondo scale $k_BT_K$.
While the phenomenology of Kondo physics is often explored with macroscopic probes of 
the low temperature quasiparticle excitations, 
electron spectroscopies detect them as the Kondo resonance near the Fermi energy 
\cite{Ehm07a,Grio97a,Tern09a}. 
Involving core levels adds element and orbital specificity
and may allow observing the effective impurity valence \cite{Fugg83a,Malt96a,Laub97a}.  

Hybridization and singlet formation being intertwined, $T_K$ sets the scales of both the change of average impurity 
orbital occupation and the screening of its effective paramagnetic moment.
As a result, the magnetic response gradually changes from being Curie-Weiss like at $T \gg T_K$ towards
a finite, Pauli-like susceptibility as $T \rightarrow 0$.
This generic behavior is well accounted for by the single impurity Anderson and Kondo models
and remains essentially valid at elevated impurity density as long as $T \gtrsim T_K$.
In Kondo lattices the coherence temperature $T^*$ emerges as a yet lower energy scale. 
It characterizes the onset of  in\-ter\-ac\-tions between impurities, 
which ultimately lead to the formation of coherent heavy-fermion bands at $T \ll T^*$
\cite{Hews93a}. 

Hybridization concurrently induces indirect magnetic coupling between the localized moments,
leading to a competition between the tendencies to screen the local moments on one hand and to order finite moments on the other
\cite{Doni77a,Hews93a,Coqb05a}. 
It is therefore interesting to specifically observe the local magnetic moments in dense Kondo systems
across a temperature window ranging from well above to well below the Kondo temperature.
Element and orbital specificity in probing magnetic response can be provided by means of spectroscopy, 
i.~e.~x-ray absorption (XA) and magnetic circular dichroism (XMCD) 
\cite{Schu07b}.
XMCD determines the magnetic polarization of
the impurity orbitals and hence provides an adequate mea\-sure of local susceptibility.
We demonstrate below that XMCD measurements on a sufficiently well defined Kondo lattice system allows addressing all
essential ingredients governing magnetic behavior, i.~e.~crystal field (CF) splitting, Kondo screening
and magnetic coupling.

Previous attempts at observing the magnetic response of Ce based Kondo or 
heavy fermion materials by XMCD appear to have been of limited success.
In all instances the Ce moments indicated Curie-Weiss like $4f$ behavior,
even where bulk magnetization clearly showed a transition to a Pauli type paramagnetism
\cite{Naka06a,Miya01a,Miya02a,Miya04a,Shio03a,Shio03b}.
Various mechanisms have been proposed to account for the apparent failure of XMCD
to reproduce the macroscopic magnetic characteristics.
Our results below indicate that the surface sensitivity of electron yield measurements
is a likely cause of the observed discrepancies.

For the present work we have chosen an approach where the relevant physics is confined
to a thin layer at specimen surfaces, ideally matching the probing depth of XMCD.
For the specific system in question, ordered CePt$_5$ surface alloys prepared on Pt(111) substrates,
the Kondo resonance \cite{Andr95a,Garn97a,Garn98a} 
and even signatures of incipient coherence \cite{Klei11a}
have been observed previously.


As established in previous work  \cite{Tang93a,Badd97a,Esse09a} 
thin Ce-Pt surface alloy films 
were prepared in situ by evaporation of Ce onto clean Pt(111) and subsequent annealing to $\approx 900$ K. 
The alloy film thickness was controlled by the amount of deposited Ce. 
Hexagonal LEED patterns indicated 2$\times$2 superstructures with thickness dependent orientation
and partial relaxation of the lateral tensile strain exerced by the substrate, qualitatively well in accordance with the literature.
Here we primarily present and discuss results obtained for an alloy layer containing $\approx$4 hexagonal atomic CePt$_2$
planes (equiv.~to Ref.~\onlinecite{Klei11a}). 
For comparison, we also use data collected at 30\% of this initial Ce coverage,   
a non-rotated hexagonal 2$\times$2 alloy with reduced sharpness of the electron diffraction spots.

\begin{figure}[]
\epsfig{file=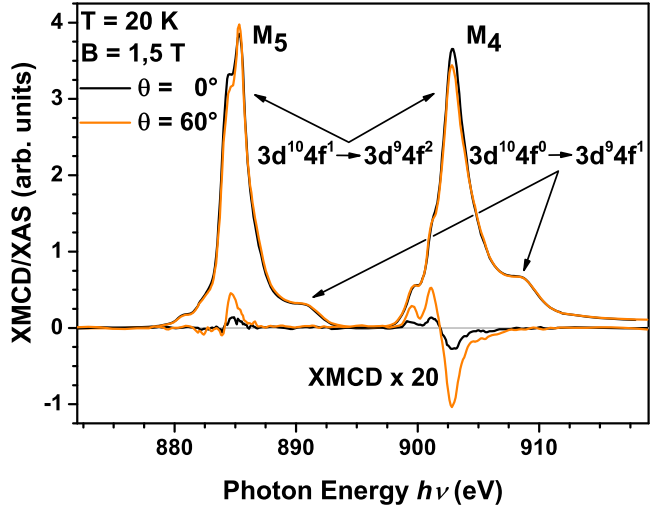,width=8.6cm}
\caption{\label{fig1:specdic} Normal and oblique incidence Ce M$_{4,5}$ XA and XMCD spectra  (average and difference, respectively, of data   taken at $\mu_0H=\pm 1.5$ T) of 4 u.~c.~CePt$_5$/Pt(111).}
\end{figure}

Ce M$_{4,5}$ XA and XMCD spectra were acquired in the total electron yield mode 
(TEY) with a custom ultrahigh vacuum superconducting magnet apparatus, 
attached to the PM3 bending magnet beamline at the BESSY II synchrotron radiation facility in Berlin, Germany. 
XMCD was measured at constant helicity ($p\approx$ 0.93) and by alternating the applied field 
($\mu_0H = \pm 1.5$ T)
at e\-ve\-ry photon energy setting. 
Magnetic anisotropy was probed by ta\-king normal 
($\theta=0^\circ$) 
and off normal incidence ($\theta=60^\circ$) spectra in the temperature range 13~K~$<T< 270$~K. 
The XMCD response consistently proved linear in the applied field.
We hence take the Ce polarization at finite field 
as a measure of the local paramagnetic $4f$ susceptibility. 
Total Ce $4f$ moments $\mu_{Ce}$ are derived from the orbital moment sum rule
\cite{Thol92a} assuming
the atomic value $g=6/7$ for the Land{\'{e}} factor and hence $\mu_{Ce} = 0.75 \mu_L$.
This procedure is most appropriate since
the $\mu_L$ sum rule is robust in the light rare earth metals \cite{Vand98a,Schi94a},
whereas proper evaluation of the spin sum rule
is impeded by strong configuration interaction between M$_5$ and M$_4$
excitations \cite{Thol85a,Schi94a,Tera96a}.
Still, XMCD sum rule evaluations require the TEY to be linear in the XA cross section,
which critically depends on the ratio of absorption length and electron escape depth $\lambda_e$
\cite{Naka99a,Abba92a}.
Beyond the material presented below, we have therefore collected auxiliary angle and alloy thickness de\-pen\-dent
datasets from which we determine  $\lambda_e \approx 1.5 \dots 2$ nm and 
conclude that TEY saturation induced relative errors in the magnetic moment evaluations amount to less than $5\%$.
The experimental $\lambda_e$, while comparable to the late $3d$ metals \cite{Naka99a},
 is considerably smaller than frequently assumed for rare earth materials \cite{Naka06a,Abba92a}.
XMCD experiments with TEY detection thus require particular caution,
especially when a comparison to bulk properties is aimed at \cite{Dall04a}.
A more detailed account of thickness dependent XA and XMCD results shall be given elsewhere 
\cite{Praeunp}. 


Fig. \ref{fig1:specdic} shows normal and grazing incidence Ce M$_{4,5}$ XA   
and XMCD spectra of the CePt$_5$ film at $T=20$ K. 
The XA spectra are composed of a multiplet and a high energy shoulder 
corresponding to core excited states of $f^2$ and $f^1$ character, respectively 
\cite{Fugg83a,Thol85a,Delo99a,Dall04a}. 
The appearance of the latter in the XA spectra  
reflect an $f^0$ admixture in the initial state valence configuration. 
The Kondo interaction entails a temperature dependent ${4f}$ occupation $n_{4f}(T/T_K)$
\cite{Bick87b,Hews93a}
and relative $f^0 \rightarrow f^1$ spectral weights can be used to estimate $T_K$ 
\cite{Delo99a,Roth99a}. 

\begin{figure}[]
\epsfig{file=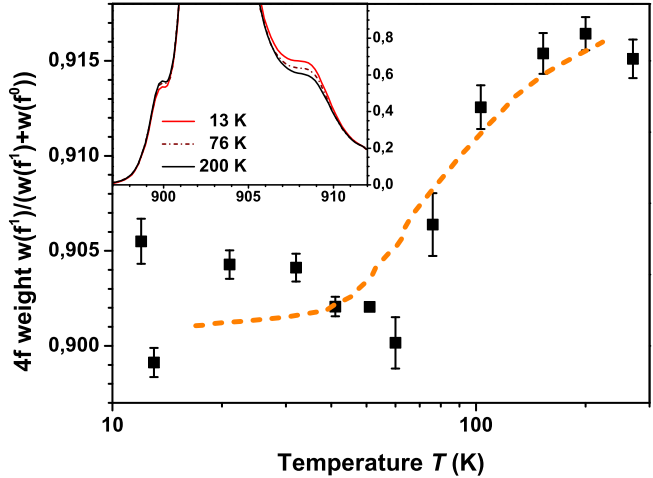,width=8.6cm}
\caption{\label{fig2:nfvsT} Evolution of $n_{4f}$ with temperature, evaluated from $f^1$ and $f^0$ initial state
spectral weights (symbols). Single impurity $n_{4f}$ in non-crossing approx., extracted
from Ref.~\onlinecite{Bick87b} (rescaled, dashed line). Inset: detail of Ce M$_4$ XA spectra, highlighting
the temperature dependent $f^0$ spectral weight.}
\end{figure}

Due to incomplete knowledge about, e.~g., initial and excited state hybridization,
the determination of $n_{4f}$ from XA experiments is not fully quantitative.
Comparing the maximum relative changes of the $f^1$ and $f^0$ initial state spectral weights
we estimate that $n_{4f} \approx 0.9$ in the low temperature limit.
On this basis we plot $n_{4f}$ vs.~specimen temperature in Fig.~\ref{fig2:nfvsT}.
Along with the experimental data, we show rescaled calculated results for the single impurity 
Kondo problem  from Ref.~\onlinecite{Bick87b}. 
The good agreement reflects the well-known observation that the high temperature ($T \gtrsim T_K$)
behaviour of heavy fermion systems resembles the one of single impurities
\cite{Hews93a}.
Based on this comparison, we estimate $T_K \approx 85$ K, fully in line with the observation
of the Kondo resonance at $T=66$ K in photoemission.
\cite{Klei11a}.

Returning to the XA and XMCD data in Fig.~\ref{fig1:specdic}, we observe minor spectral differences 
between normal and off normal incidence spectra, i.~e.~small linear dichroism owing to the hexagonal symmetry. 
By contrast, the magnetic anisotropy  derived from the 
XMCD magnitude is much stronger and amounts to about  $\chi_{60^\circ} / \chi_{0^\circ} = 4$ at $T=20$ K, 
which translates to an anisotropy 
$\chi_{\|} / \chi_\perp = 5$ between the susceptibilities along and perpendicular to the hexagonal
axis, respectively.
The  $m_J$ levels of the $4f^1$ 
configuration not being mixed by a hexagonal CF,
we can unanimously determine the ground level from
the single ion anisotropy.
Based on second order expressions for the susceptibility \cite{Luek79a} we write the anisotropy as
\begin{displaymath}\label{anisFormula}
\frac{\chi_{\|}}{\chi_\perp}= 
                         \frac{ 9 + \frac{16k_BT}{\Delta_1}
                       - \left ( \frac{16k_BT}{\Delta_1} - \frac{10k_BT}{\Delta_2-\Delta_1} \right ) e^{\frac{-\Delta_1}{k_BT}}
                       - \frac{10k_BT}{\Delta_2-\Delta_1}e^{\frac{-\Delta_2}{k_BT}}}{1 + 9e^{\frac{-\Delta_1}{k_BT}} + 25e^{\frac{-\Delta_2}{k_BT}}}
\end{displaymath}
\noindent which only depends on temperature and the CF splitting, given by $\Delta_1 = E_{\rm{3/2}}-E_{\rm{1/2}}$
and $\Delta_2 = E_{\rm{5/2}}-E_{\rm{1/2}}$.
It follows that the observed magnetic anisotropy cannot be obtained unless $m_J=1/2$ is the ground level,
in line with the analysis of polycrystalline CePt$_5$ in Ref.~\onlinecite{Luek79a}.

In the limit of $k_BT \ll \Delta_1,\Delta_2$
and in absence of magnetic coupling between Ce sites,
the paramagnetic response is expected to follow a Curie law 
with a  reduced $4f$ moment compared to the free ion value ($2.54 \mu_B$). 
For the  $m_J=1/2$ ground level one obtains
$\mu_{1/2}=g\sqrt{3/4}\mu_B$ along the hexagonal axis.
To additionally allow for Kondo screening we introduce an effective moment,
related to the local susceptibility via $\chi_{loc} = {\mu_{eff}^2}/{3k_BT} $.


\begin{figure}[]
\epsfig{file=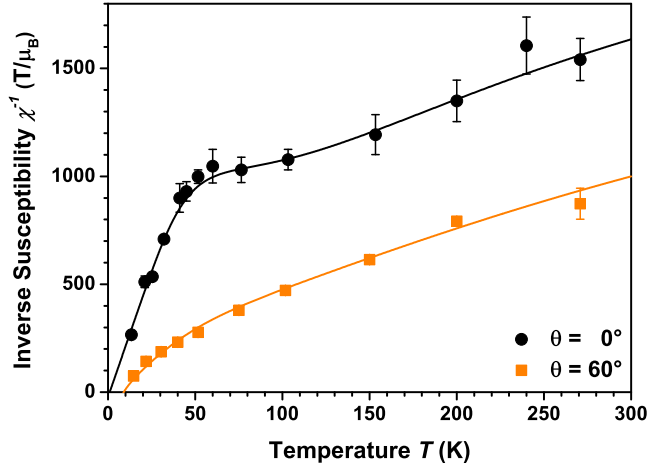,width=8.6cm}
\caption{\label{fig3:InvChivsT} Symbols: normal and off-normal incidence inverse local Ce $4f$ susceptibility, derived from XMCD. Lines: Curie-Weiss fits to $\chi^{-1}_{0^\circ,60^\circ}$
with $\Delta_{1,2}$ as given in the text.}
\end{figure}

Figure \ref{fig3:InvChivsT} plots the inverse local Ce $4f$ susceptibility vs.~temperature in the range of 13 K $\leq T \leq 270$ K. 
The pronounced slope changes in the normal incidence $\chi^{-1}_{\|}(T)$ are due to the CF level structure and 
are adequately accounted for with  $\Delta_1= 16\pm 1$ meV and $\Delta_2= 91\pm 15$ meV. 
 We note that $\Delta_1$ is significantly smaller than the corresponding value (28 meV) in bulk CePt$_5$ 
\cite{Luek79a}
as a consequence of the dilated CePt$_5$ lattice when grown on Pt(111).
Thickness dependent measurements at different levels of strain relaxation confirm this assignment
\cite{Praeunp}. 
The low temperature magnetic response obviously follows a Curie-Weiss law 
with positive paramagnetic Curie-Weiss temperature $\Theta_{p}$, 
indicative of ferromagnetic correlations in the alloy film.
The angle dependence of $\Theta_{p}$ derives from the anisotropic effective moment. 
Allowing a Weiss mean field we write the inverse local susceptibility as 
$\chi^{-1}_{loc}={3k_BT}/{\mu_{eff}^2} + \lambda$ 
and see that $\mu_{eff}$ can be evaluated at low temperature from the slope of $\chi^{-1}_{loc}$, 
irrespective of the coupling strength $\lambda$.

The low temperature $4f$ moment resulting from the $\chi_\|^{-1}$ data of Fig.~\ref{fig3:InvChivsT} 
amounts to $\mu_{eff} = 0.43(2) \mu_B$, 
and is thus reduced to below $60\%$ of the expected value. 
We take this as immediate evidence of Kondo screening of the local Ce moment.
The essentially temperature independent XMCD spectral shape clearly indicates
that spin and orbital parts of the Ce $4f$ moment are screened alike,
as expected from the Coqblin-Schrieffer model \cite{Coqb69a}.
Calculated scenarios do report on relative moment reductions 
of the same order
upon cooling through the Kondo temperature 
\cite{Beac08a,Li10a}.
In our experiments we thus directly observe partially screened Ce $4f$ moments with
mutual interactions of predominantly ferromagnetic character .

These findings appear to contrast the observation of antiferromagnetism in bulk CePt${_5}$ at very low temperature
($T_N = 1$ K) \cite{Schr88a}.
We suggest that our observations result from the specimen geometry: in the ultrathin
surface alloys, the magnetic fluctuations may be dominated by interactions within the atomic layers
hosting the Ce atoms, while interlayer interactions might become important at lower temperatures and with a larger number of layers, only.
Concerning ordering within the hexagonal CePt$_2$ planes, our data suggest a possible ferromagnetic instability
at $T \lesssim 10$ K.
Accordingly, we conjecture that the volume phase of CePt$_5$ probably is a layer wise (type II) antiferromagnet. 

\begin{figure}[]
\epsfig{file=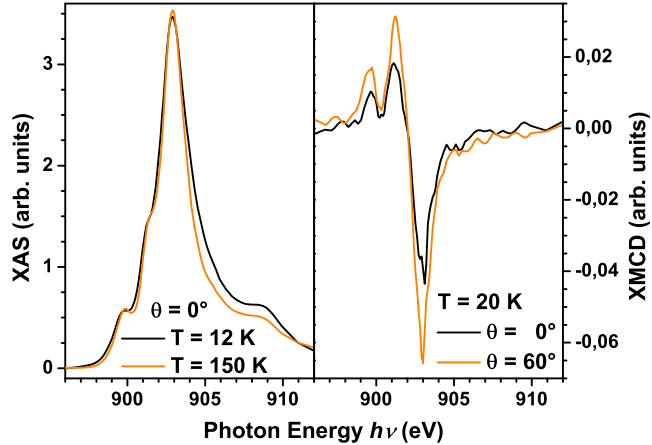,width=8.6cm}
\caption{\label{fig4:thinsample} 
a) Ce M$_4$ XA at low thickness with increased $f^0$ weight at low temperature.
b) Normal and off-normal Ce M$_4$ XMCD at 20 K with reduced anisotropy compared to 
Fig.~\ref{fig1:specdic}.}
\end{figure}

The numerical results for the CF energies $\Delta_1, \Delta_2$ call for a cross check of the magnitude
of the magnetic anisotropy on the basis of the formula given above.
Indeed, a larger anisotropy is expected with these CF level splittings,
further enhanced by the magnetic coupling ($\Theta_p>0$).
Given the surface sensitivity of XMCD one might suspect the occurrence of a modified
magnetic anisotropy at the surface.
We have therefore investigated an alloy film of reduced thickness for comparison,
as outlined above.
XA spectra at high and low temperature, respectively, show  magnitudes of $f{^0}$ initial state character
and valence change upon cooling which are comparable to the data presented in Figs.~\ref{fig1:specdic} and \ref{fig2:nfvsT}, suggesting a similar Kondo screening to occur.
The XMCD response, however, is significantly altered (Fig.~\ref{fig4:thinsample}), simultaneously being enhanced at normal incidence and reduced in off-normal geometry.
The resulting magnetic anisotropy is considerably reduced ($\chi_{60^\circ} / \chi_{0^\circ} = 1.6$ at $20$ K) 
compared to the thicker alloy film.
The assumption of the subsurface CePt$_2$ layer possessing a  strongly reduced anisotropy  and a homogeneous CePt$_5$ film
underneath provides a consistent scenario for our observation and proves to be in accordance with results on a wider range of surface alloy thickness \cite{Praeunp}.
Such a modification can readily be rationalized as to arise from the different coordination of the Ce atoms in the topmost Ce-containing layer.
In particular, a strong reduction of $\Delta_1$ would provide a rational basis for reduced magnetic anisotropy of the surface-near Ce ions.
These considerations strongly underline that great caution is required when applying surface sensitive methods
to rare earth alloys.
In the present case, the discrimination of surface vs.~``bulk'' properties is facilitated by the fact that 
reasonably well ordered alloy films can be prepared over an extended range of alloy thickness.

We finally note that  the Ce $4f$ susceptibility does not bend off
to a finite, temperature independent value.
Thus, while impurity Kondo screening is evident in our experimental data, there is no obvious
signature of coherence for $T \geq 13$ K, suggesting
 a lower scale for the full development of the heavy fermion state.
In addition, the proximity to a ferromagnetic instability may  stabilize the Curie-Weiss
behavior in the CePt$_5$ surface alloys.

In conclusion, we have presented an experimental investigation of the magnetic response of a Kondo lattice
by examining the  local Ce $4f$ susceptibility in ordered CePt$_5$ surface alloys with XMCD.
Thin film Kondo lattices prove an ideal match to the surface sensitivity of XMCD in the TEY mode.
Varying the alloy thickness allows assessing the importance of 
surface induced modifications to the magnetic properties.
While the anisotropic paramagnetic response allows characterizing the hexagonal CF, 
the quantitative XMCD sum rule evaluation reveals the Kondo-screened nature of the Ce moments.
Our finding of a positive Curie Weiss temperature suggests a ferromagnetic instability to occur around $T \lesssim 10$ K,
presumably due to RKKY coupling within the hexagonal CePt$_2$ planes.
It will be most interesting to carry through these investigations to lower temperatures in order
to follow the evolution to a potentially magnetically ordered coherent heavy fermion state.
Due to its inherent element and orbital specificity,  direct evidence 
of both ordered and screened local moments would be provided by XMCD. 
Work along these lines is currently in progress.

\begin{acknowledgments}
We gratefully acknowledge the support by the BESSY staff, R. Follath, T.~Kachel and H.~Pfau in particular.
This work was financially supported by the Deutsche Forschungsgemeinschaft within research
unit FOR1162. Additional partial funding through BMBF under contract no.~{05ES3XBA/5}
is gratefully acknowledged.
\end{acknowledgments}


\end{document}